\documentclass{prep}

\usepackage{graphicx,amsmath}
\usepackage{pslatex}
\usepackage{natbib}

\newcommand{\gsim}{\stackrel{>}{{}_\sim}}
\newcommand{\lsim}{\stackrel{<}{{}_\sim}}

\title{On the Viscosity of Emulsions}

\author{Klaus Kroy\thanks{klaus@pmmh.espci.fr}, Isabelle Capron,
  Madeleine Djabourov} 

\affiliation{Physiqe Thermique, ESPCI. 10, rue Vauquelin, Paris.
  France}

\begin{document}

\maketitle

\begin{abstract}
  Combining direct computations with invariance arguments, Taylor's
  constitutive equation for an emulsion can be extrapolated to high
  shear rates.  We show that the resulting expression is consistent
  with the rigorous limits of small drop deformation and that it bears
  a strong similarity to an \emph{a priori} unrelated rheological
  quantity, namely the dynamic (frequency dependent) linear shear
  response.  More precisely, within a large parameter region the
  nonlinear steady--state shear viscosity is obtained from the real
  part of the complex dynamic viscosity, while the first normal stress
  difference is obtained from its imaginary part. Our experiments with
  a droplet phase of a binary polymer solution (alginate/caseinate)
  can be interpreted by an emulsion analogy. They indicate that the
  predicted similarity rule generalizes to the case of moderately
  viscoelastic constituents that obey the Cox--Merz rule.
\end{abstract}

\section{Introduction}
Apart from their technological importance, emulsions have served as
model systems accessible to rigorous theoretical modeling. The study
of emulsions consisting of droplets of a liquid dispersed in another
liquid has thus contributed substantially to our understanding of the
rheology of complex fluids. However, although major theoretical
achievements date back to the beginning of the $20^{\rm th}$ century,
further progress turned out to be difficult.  The macroscopic
rheological properties of emulsions are determined by the reaction of
the individual drops to the flow field, which in turn is modified by
the presence of other drops. The mutual hydrodynamic interactions of
drops complicates substantially the mathematical description.
Moreover, depending on system parameters and flow type, droplets may
break under steady flow conditions if a certain critical strain rate
is exceeded.  Rigorous calculations of the constitutive equation have
therefore concentrated on very dilute emulsions and on conditions
where drops are only weakly deformed.  Sometimes, however, it is
desirable to have an approximate expression, which --- though not
rigorous --- can serve for practical purposes as a quantitative
description in the parameter region beyond the ideal limits. As far as
the dependence of the viscoelastic properties of an emulsion on the
volume fraction $\phi$ of the dispersed phase is concerned, such an
approximation has been given by \cite{oldroyd:53}.  It is not rigorous
beyond first order in $\phi$ but serves well some practical purposes
even at rather high volume fractions.  It seems not reasonable to look
for a comparably simple approximation for the dependence of shear
viscosity $\eta$ on shear rate $\dot \gamma$ that covers the whole
range of parameters, where all kinds of difficult break--up scenarios
are known to occur.  In the next section, we propose instead a less
predictive expression which contains an average drop size $R$ (that
may change with shear rate) as a phenomenological parameter. The
latter has to be determined independently either from theory or
experiment. It turns out, however, that for a substantial range of
viscosity ratios and shear rates, the expression for $\eta (\dot
\gamma)$ is to a large extent independent of morphology.  For
conditions, where drops do not break outside this region, we point out
a similarity relation between this expression and the frequency
dependent viscoelastic moduli $G' (\omega)$, $G'' (\omega)$, similar
to the \emph{Cox--Merz} rule in polymer physics.  More precisely, we
show that in the limit of small drop deformation, the constitutive
equation of an emulsion composed of Newtonian constituents of equal
density can be obtained from the frequency dependent linear response
to leading order in capillary number ${\cal C}$ and (reciprocal)
viscosity ratio $\lambda^{-1}$.  And we argue that the identification
is likely to represent a good approximation beyond this limit in a
larger part of the ${\cal C}-\lambda$ parameter plane.
Experimentally, this similarity relation can be tested directly,
without interference of theoretical modeling, by comparing two
independent sets of rheological data. In summary, our theoretical
discussion provokes two major empirical questions. (1) If drops break:
does the expression for the viscosity derived in
Eq.(\ref{eq:cm_miracle}) describe the data with $R$ the average drop
size at a given shear rate? (2) If drops do not break below a certain
characteristic capillary number $C^*(\lambda)$: does the proposed
similarity rule hold?  To what extent does it generalize to
non--Newtonian constituents? In Section~\ref{sec:experiment} we
address mainly the second question by experiments with a quasi--static
droplet phase of a mixture of moderately viscoelastic polymer
solutions.

\section{Theory}\label{sec:theory}
\subsection{Taylor's constitutive equation for emulsions} 
A common way to characterize the rheological properties of complex
fluids such as emulsions, suspensions, and polymer solutions, is by
means of a \emph{constitutive equation} or an equation of state that
relates the components $p_{ij}+p\delta_{ij}$ of the \emph{stress
  tensor} to the \emph{rate--of--strain tensor} $e_{ij}$.  This
relation can account for all the internal heterogeneity and the
complexity and interactions of the constituents if only the system may
be represented as a homogeneous fluid on macroscopic scales.  The form
of possible constitutive equations is restricted by general
\emph{symmetry arguments}, which provide guidelines for the
construction of phenomenological expressions
\cite[]{oldroyd:50,oldroyd:58}. On the other hand, for special model
systems the constitutive equations may be calculated directly at least
for some restricted range of parameters. An early example for a direct
computation of the constitutive equation of a complex fluid is
Einstein's formula
\begin{equation}
  \label{eq:einstein}
  \eta=\eta_c\left(1+\frac52\phi\right) 
\end{equation}
for the shear viscosity $\eta\equiv p_{12}/e_{12}$ of a dilute
suspension (particle volume fraction $\phi\ll1$). It is obtained by
solving Stokes' equation for an infinite homogeneous fluid of
viscosity $\eta_c$ containing a single solid sphere.  For a
sufficiently dilute suspension, the contributions of different
particles to the overall viscosity $\eta$ can be added independently,
giving an effect proportional to $\phi$.  In close analogy
\cite{taylor:32} calculated $\eta$ for a steadily sheared dilute
suspension of droplets of an incompressible liquid of viscosity
$\eta_d\equiv \lambda \eta_c$ in another incompressible liquid of
viscosity $\eta_c$. For weakly deformed drops he obtained
\begin{equation}
  \label{eq:taylor}
  \eta= \eta_c\left(1+\phi\frac{5\lambda+2}{2\lambda +2}\right) \equiv
  \eta_T \;,
\end{equation}
which we abbreviate by $\eta_T$ in the following.  This expression
includes Einstein's result as the limiting case of a highly viscous
droplet, $\lambda\to\infty$. As in Einstein's calculation,
interactions of the drops are neglected.  The result is independent of
surface tension $\sigma$, shear rate $\dot \gamma$, and drop radius
$R$; i.e., it is a mere consequence of the presence of a certain
amount $\phi$ of dispersed drops, regardless of drop size and
deformation (as long as the latter is small). Moreover, the dynamic
(frequency dependent) linear response of an emulsion has been
calculated by \cite{oldroyd:53}. His results are quoted in
section~\ref{sec:cox-merz} below.

Under steady flow conditions, drop deformation itself is proportional
to the magnitude of the rate--of--strain tensor $e_{ij}$.  More
precisely, for simple shear flow with constant shear rate $\dot
\gamma$, the characteristic measure of drop deformation
for given $\lambda$ is the \emph{capillary number}
\begin{equation}
  \label{eq:capillary_number}
  {\cal C}=\frac{\eta_c R\dot\gamma}{\sigma}  \;,
\end{equation} 
also introduced by \cite{taylor:34}.  It appears as dimensionless
expansion parameter in a perturbation series of the drop shape
under shear. To derive Taylor's Eq.(\ref{eq:taylor}) it is sufficient
to represent the drops by their spherical equilibrium shape.  Aiming
to improve the constitutive equation,
\cite*{schowalter-chaffey-brenner:68} took into account deformations
of drops to first order in $\cal C$. The refined analysis did not
affect the off--diagonal elements of the constitutive equation, i.e.\ 
Taylor's Eq.(\ref{eq:taylor}) for the viscosity, but it gave the
(unequal) normal stresses to order $O({\cal C}\dot\gamma)$.  Another
limit, where exact results can be obtained, is the limit of large
viscosity ratios $\lambda\to\infty$
\cite[]{frankel-acrivos:70,rallison:80}.  To clarify the physical
significance of the different limits we want to give a brief
qualitative description of the behavior of a suspended drop under
shear, based on work by \cite{oldroyd:53} and \cite{rallison:80}.

In a quiescent matrix fluid of viscosity $\eta_c$, a single weakly
deformed drop relaxes exponentially into its spherical equilibrium
shape; i.e., defining dimensionless deformation by $D:=(a-b)/(a+b)$
with $a$ and $b$ the major and minor axis of the elongated drop, one
has for a small initial deformation $D_0$,
\begin{equation}
  \label{eq:d_relax}
   D=D_0\, \text{e}^{-t/\tau_1}\;.
\end{equation}
The characteristic \emph{relaxation time} \cite[]{oldroyd:53}
\begin{equation}
  \label{eq:relaxation}
  \tau_1 = \frac{\eta_cR}{\sigma}
  \frac{(2\lambda + 3) (19\lambda + 16)}{40 (\lambda + 1)} 
\end{equation}
also characterizes the macroscopic \emph{stress relaxation} in an
unstrained region of a dilute emulsion. At $\omega\tau_1\simeq 1$ one
observes the characteristic relaxation mode in the frequency dependent
moduli. The relaxation time diverges for $\lambda/\sigma\to\infty$
since it takes longer for a weak surface tension to drive a viscous
drop back to equilibrium.  What happens if the matrix is steadily
sheared at shear rate $\dot\gamma$?  For $\dot\gamma \tau_1 \ll 1$,
the flow induced in the drop by the external driving is weak compared
to the internal relaxation dynamics and the equilibrium state is only
slightly disturbed, i.e., the drop is only weakly deformed.
Similarly, for large viscosity ratio $\lambda$, the elongation of the
drop becomes very slow compared to vorticity, and hence again very
small in the steady state, even if $\tau_1\dot\gamma$ is not small.
Technically, this is due to the asymptotic proportionality to
$\lambda^{-1}$ of the shear rate within the drop. In both limits of
weak deformation, the time $\tau_1$ also controls the orientation of
the major axis of the drop with respect to the flow according to
\begin{equation}
  \label{eq:angle}
   \frac{\pi}{4}-\frac12 \arctan(\tau_1\dot\gamma ) \;.
\end{equation}
Eq.(\ref{eq:d_relax}) and Eq.(\ref{eq:angle}) both can be used to
determine the surface tension $\sigma$ from observations of single
drops under a microscope.  In passing, we note that the classical
method based on the result obtained by \cite{taylor:34} for the
steady--state deformation, can only be used if $\lambda$ is not too
large, whereas Eq.(\ref{eq:d_relax}) and Eq.(\ref{eq:angle}) are more
general.

The exact calculations mentioned so far became feasible because (and
are applicable if) deviations of the drop from its spherical
equilibrium shape are small.  On the other hand, if neither the
capillary number (or $\tau_1\dot\gamma$) is small nor the viscosity
ratio is large, i.e.,
\begin{equation}
  \label{eq:restriction}
  {\cal C} \gsim 1 \qquad \text{and} \qquad \lambda\lsim 1 \;,
\end{equation}
drops can be strongly deformed by the symmetric part of the flow
field. Experiments with single drops by \cite{grace:82} and others
have shown that this eventually leads to drop break--up if $\lambda$
is smaller than some critical viscosity ratio. No general rigorous
result for the viscosity of an emulsion is known in this regime, where
arbitrary drop deformations and break--up may occur. Below we will
also be interested in such cases, where the conditions required for
the rigorous calculations are not fulfilled.

\subsection{Second--order theory}\label{sec:improving}
For the following discussion we introduce some additional notation.
The rate--of--strain tensor $e_{ij}$ and the vorticity tensor
$\omega_{ij}$ are defined as symmetric and antisymmetric parts of the
velocity gradient $\partial_jv_i$. In particular, for a steady simple
shear flow $v_i=\dot\gamma x_2\delta_{i1}$, and
\begin{equation}
  \label{eq:strain}
  \partial_jv_i =  e_{ij}+ \omega_{ij} = \dot\gamma
  (\delta_{i1}\delta_{j2}+\delta_{i2}\delta_{j1})/2
  + \dot\gamma(\delta_{i1}\delta_{j2}-\delta_{i2}\delta_{j1})/2 \;.
\end{equation}
The components $p_{ij}$ of the stress tensor in the shear plane ($i,j
\in \{1,2\}$) as obtained for finite $\lambda$ by
\cite{schowalter-chaffey-brenner:68} read
\begin{equation}
  \label{eq:schowalter}
  p_{ij}=2\eta_T e_{ij} 
  - \eta_c \phi \tau_1 
  \! \frac{19\lambda + 16}{(2\lambda + 3)(\lambda + 1)} 
  {\cal D}e_{ij}
  + \eta_c \phi \tau_1 \dot \gamma^2 \! 
  \frac{25\lambda^2+41\lambda+4}{14 (2\lambda+3)
    (\lambda+1)^2}\delta_{ij}
  + O(\dot\gamma {\cal C}^2) \;.
\end{equation}
As usual, the material derivative has been defined by
\begin{equation}
  \label{eq:material_derivative}
  {\cal D} c_{ij} := \partial_t + v_k \partial_k c_{ij} 
  + \omega_{ik} c_{kj} + \omega_{jk} c_{ik} \;,   
\end{equation}
where summation over repeated indices is implied, and the first two
terms vanish for steady shear flow. Note that since ${\cal D} e_{ij}$
is diagonal, Eq.(\ref{eq:schowalter}) implies $\eta=\eta_T+ O({\cal
  C}^2)$, and hence Eq.(\ref{eq:taylor}) remains valid to first order
in ${\cal C}$ as we mentioned already.

\begin{figure}
  \begin{center}
    \leavevmode 
     \includegraphics[width=0.5\columnwidth]{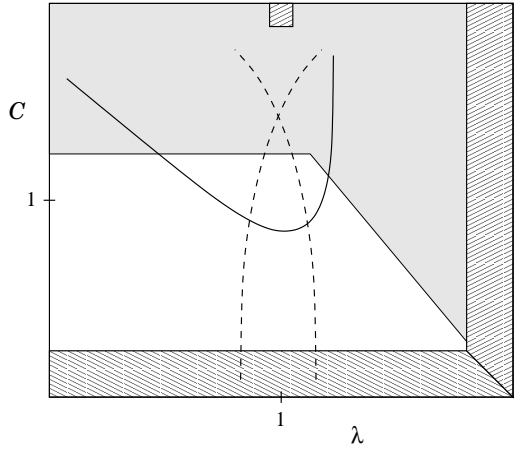}
    \caption{Schematic representation of the limits (hatched) where 
      Eqs.(\ref{eq:miracle}), (\ref{eq:cm_miracle}) for the viscosity
      $\eta$ of an emulsion give rigorous results.  In the shaded
      region they predict $\eta$ to be practically independent of
      capillary number. The frame of the box is meant to comprise the
      whole $\cal C$-$\lambda$ parameter plane from zero to infinity
      (small Reynolds number understood).  The curved solid line is a
      sketch of the break--up curve for steadily sheared isolated
      Newtonian drops according to \cite{grace:82}.  Dashed lines
      indicate schematically our viscosity measurements (see
      Section~\ref{sec:experiment}).}
    \label{fig:limits}
  \end{center}
\end{figure}

Can we extrapolate the exact second order result
Eq.(\ref{eq:schowalter}) for the stress tensor to arbitrary ${\cal C}$
and $\lambda$ by using the constraints provided by general invariance
arguments?  For example, since the shear stress has to change sign if
the direction of the shear strain is inverted whereas the normal
stresses do not, the shear stress and the normal stresses have to be
odd/even functions of $\dot \gamma$, respectively.  From this
observation we could have foreseen that Eq.(\ref{eq:taylor}) cannot be
improved by calculating the next order in $\dot \gamma$, i.e., by
considering droplet deformation to lowest order.  More important are
Galilean invariance and invariance under transformations to rotating
coordinate frames, which give rise to the material derivative
introduced above. Applying the operator $(1+\tau_1 {\cal D})$ to
Eq.(\ref{eq:schowalter}) adds to the right hand side of the equation a
term $2\eta_T\tau_1{\cal D} e_{ij}$ plus a term of order $O(\dot\gamma
{\cal C}^2)$, so that one obtains (in the shear plane)
\begin{equation}
  \label{eq:acrivos}
  p_{ij}+\tau_1 {\cal D}\,p_{ij} = 2\eta_T(e_{ij}+\tau_2
  {\cal D}e_{ij}) + \eta_c \phi \tau_1 \dot \gamma^2 \!
    \frac{25\lambda^2+41\lambda+4}{14 (2\lambda+3)
      (\lambda+1)^2}\delta_{ij} + O(\dot\gamma {\cal C}^2) \;.
\end{equation}
As another short--hand notation we have introduced a second
characteristic time $\tau_2$, which to the present level of accuracy
in $\phi$ is given by
\begin{equation}
  \label{eq:tau_2}
  \tau_2/\tau_1 = 1- \phi
    \frac{19\lambda + 16}{(2\lambda + 3)(2\lambda + 2)}  +O(\phi^2)\;.
\end{equation}
It sets the time scale for \emph{strain relaxation} in an unstressed
region and was named \emph{retardation} time by \cite{oldroyd:53}.
\cite{frankel-acrivos:70} realized that up to the partly unknown terms
of order $O(\dot\gamma{\cal C}^2)$ on the right--hand side,
Eq.(\ref{eq:acrivos}) belongs to a class of possible viscoelastic
equations of state already discussed by \cite{oldroyd:58}. Hence,
setting $O(\dot\gamma {\cal C}^2) \equiv 0$ on the right--hand side of
Eq.(\ref{eq:acrivos}), we can define a (minimal) model viscoelastic
fluid that behaves identical to the emulsion described by
Eq.(\ref{eq:schowalter}) for small shear rates.  In contrast to
Eq.(\ref{eq:acrivos}), the truncated formula for the viscosity
\begin{equation}
  \label{eq:miracle}
  \eta  = \eta_T\frac{ 1 +
 \tau_1\tau_2\dot\gamma^2} {1+(\tau_1\dot\gamma)^2} =
  \frac{\eta_c}{1+(\tau_1\dot\gamma)^2}\left[ 
    1+\phi\frac{5\lambda+2}{2\lambda+2}
    +\left(1+\phi\frac{5(\lambda-1)}{2\lambda +3} \right) 
    (\tau_1\dot\gamma)^2 \right]
\end{equation}
thus obtained has a manifestly non--perturbative form.  However, no
phenomenological parameters had to be introduced. Note that
Eq.(\ref{eq:miracle}) comprises both exactly known limits: ${\cal
C}\to0$ for fixed $\lambda$, and $\lambda\to\infty$ for arbitrary
${\cal C}$. Obviously, \cite{frankel-acrivos:70} have forgotten a term
$-25\eta_c\phi e_{ij}/2$ in their Eq.(3.6) for $p_{ij}$ in the limit
$\lambda\to\infty$ for fixed $\cal C$. If the latter is included,
Eq.(\ref{eq:miracle}) is also in accord with their
$O(\lambda^{-1})-$analysis. Moreover, Eq.(\ref{eq:miracle}) has the
proper limiting behavior for $\lambda=1$, $\sigma\to 0$, i.e.\ ${\cal
C}\to\infty$, which is an extreme case of Eq.(\ref{eq:restriction}).
Since we assume equal densities for the two phases, the two--phase
fluid actually reduces to a one--phase fluid in this degenerate case,
and the viscosity is simply $\eta_c$, independent of morphology. For
illustration, the rigorous limits of Eq.(\ref{eq:miracle}) in the
${\cal C}-\lambda$ plane are depicted graphically in
Fig.~\ref{fig:limits}.  Following \cite{grace:82}, a qualitative
break--up curve for single drops under steady shear is also sketched.
In summary, Eq.(\ref{eq:miracle}) is correct for arbitrary $\lambda$
if ${\cal C}\to 0$, and for arbitrary ${\cal C}$ if
$\lambda\to\infty$, and for small and large ${\cal C}$ if $\lambda=1$.
Therefore, one can expect that Eq.(\ref{eq:miracle}) works reasonably
well within a large parameter range (small Reynolds number
understood).  This is further supported by the observation that the
error made in going from Eq.(\ref{eq:schowalter}) to
Eq.(\ref{eq:miracle}) rather concerns the \emph{shape} of the droplet
than its \emph{extension} (it consists in truncating a perturbation
series in shape parametrisation). The final result, though sensitive
to the latter, is probably less sensitive to the former. Nevertheless,
one would not be surprised to see deviations from
Eq.(\ref{eq:miracle}) when drops become extremely elongated.  Finally,
due to changes in morphology by break--up and coalescence, the avarage
drop size $R$ may change. Observe, however, that for most viscosity
ratios ($\lambda$ not close to unity), Eq.(\ref{eq:miracle}) is
practically independent of capillary number (and thus of $R$) when
break--up might be expected according \cite{grace:82} and others.  As
an analytic function that is physically known to be bounded from above
and from below (the latter at least by the viscosity of a stratified
two--phase fluid depicted in Fig.~\ref{fig:mixing}),
$\eta(\lambda,{\cal C})$ has to have vanishing slope in the ${\cal
C}-$direction for large ${\cal C}$.  According to
Eq.(\ref{eq:miracle}), $\eta(\lambda,{\cal C})$ is almost independent
of ${\cal C}$ for
\begin{equation}
  \label{eq:region}
  {\cal C} \gg {\cal C}^* \approx 
  \frac{40 (\lambda + 1)}{\sqrt3 (2\lambda + 3) (19\lambda + 16)}  \;,
\end{equation}
where ${\cal C}^*$ is the turning point in the dilute limit,
determined by $\tau_1\dot \gamma = 1/\sqrt3$.  For finite volume
fractions $\tilde\tau_1$ from Eq.(\ref{eq:tau_1_oldroyd}) replaces
$\tau_1$.  Hence, for ${\cal C} \gg {\cal C}^*$,
Eq.(\ref{eq:miracle}) and its extension to higher volume fractions
derived below in Eq.(\ref{eq:cm_miracle}) are practically independent
of drop deformation and morphology. For finite volume fractions a
rough interpolation for ${\cal C}^*$ derived from
Eq.(\ref{eq:cm_miracle}) is given by $2.4/(5(1+\phi)+4\lambda)$. In a
large part of parameter space we thus expect Eqs.(\ref{eq:miracle}),
(\ref{eq:cm_miracle}) to be applicable to monotonic shear histories
with $R$ given by the average initial radius of the droplets. For
non--monotonic shear histories, there can of course be hysteresis
effects in $\eta (\dot\gamma)$ that result from morphological changes
for ${\cal C} \gg {\cal C}^*$. These can only be avoided by
substituting for $R$ the radius corresponding to the actual average
drop size at the applied shear rate $\dot\gamma$.

\begin{figure}
  \begin{center}
    \leavevmode 
     \includegraphics[width=0.4\columnwidth]{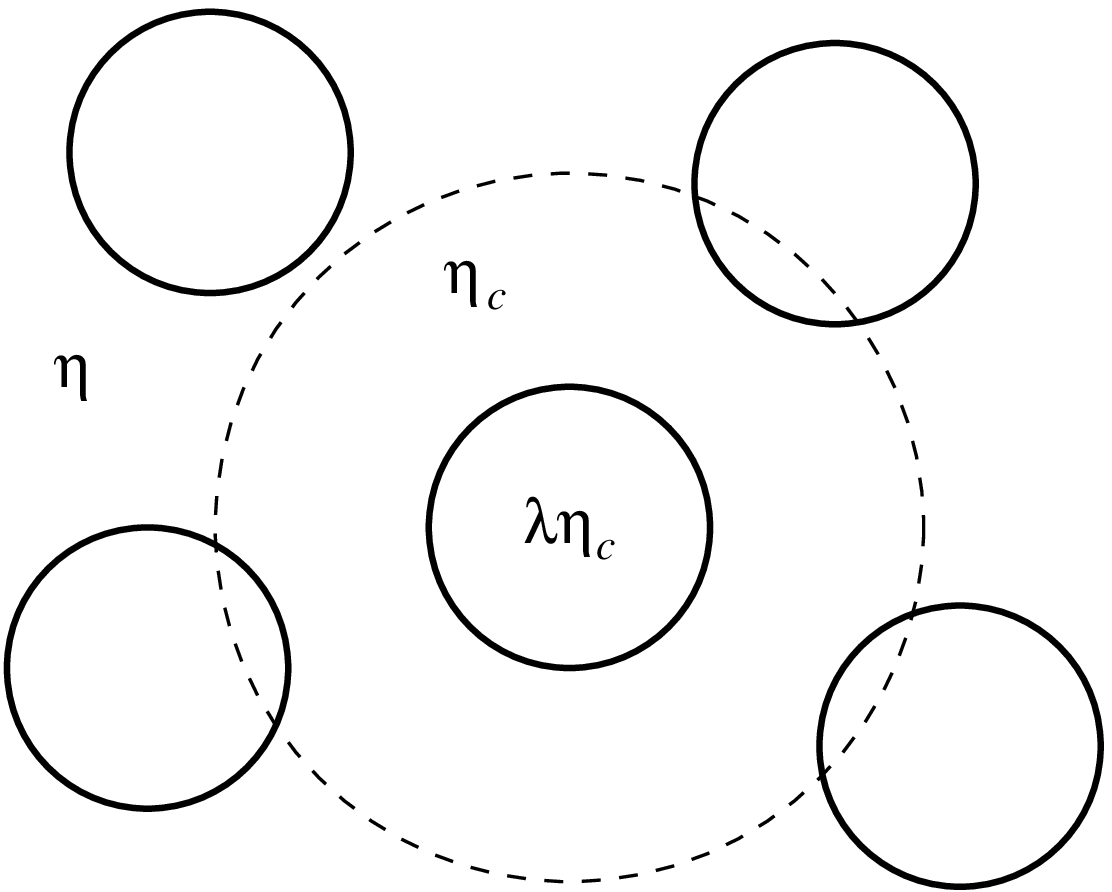}
    \caption{ Self--consistent mean--field description of
      volume--fraction effects in disordered emulsions and
      suspensions.  The fluid surrounding a test droplet is assumed to
      have the viscosity $\eta_c$ of the continuous phase/the
      viscosity $\eta$ of the whole emulsion, within/outside the
      ``free--volume--sphere'' of radius $R/\phi^{1/3}$.}
    \label{fig:mf-trick}
  \end{center}
\end{figure}

\subsection{Finite volume fractions}\label{sec:volume_fraction}
\begin{figure}
  \begin{center}
    \leavevmode 
    \includegraphics[width=0.55\columnwidth]{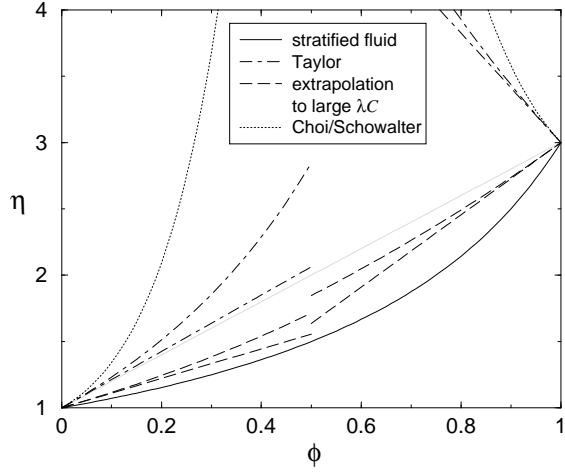}
    \caption{Comparison of different mixing rules for
      emulsions with viscosity ratio $\lambda=3$ (chosen arbitrarily).
      The dot--dashed and dashed \emph{straight} lines pertain to
      dilute emulsions described by the extrapolation formula
      Eq.(\ref{eq:miracle}), which reduces to Taylor's formula
      Eq.(\ref{eq:taylor}) for small capillary numbers. The
      corresponding \emph{curved} lines are obtained from
      Eq.(\ref{eq:cm_miracle}) where interactions of the droplets are
      taken into account in a mean--field approximation. The curved
      \emph{dotted} lines are the predictions of the cell model by
      \cite{choi-schowalter:75} for small shear rates,
      Eq.(\ref{eq:choi}).  The curved \emph{solid} line represents the
      viscosity $\eta=\eta_c/[\phi+(1-\phi)\lambda^{-1}]$ of a
      two--phase stratified fluid and is a lower bound for any
      viscosity mixing rule.}
    \label{fig:mixing}
  \end{center}
\end{figure}

A general limitation of the equations discussed so far, is the
restriction to small volume fractions. Above, we have implicitly
assumed that second order effects from drop interactions are small
compared to second order effects from drop deformation. Any direct
(coalescence) and indirect (hydrodynamic) interactions of droplets
have been neglected in the derivation. Hydrodynamic interactions can
approximately be taken into account by various types of cell models.
Recently \cite{palierne:90} proposed a self--consistent method
analogous to the Clausius--Mossotti or Lorentz--sphere method of
electrostatics. For the case of a disordered spatial distribution of
drops his results reduce to those already obtained by
\cite{oldroyd:53}.  Oldroyd artificially divides the volume around a
droplet of viscosity $\eta_d \equiv \lambda\eta_c$ into an interior
``free volume'' with a viscosity $\eta_c$ of the bare continuous phase
and an exterior part with the viscosity $\eta$ of the whole emulsion
(see Fig.~\ref{fig:mf-trick}).  According to this scheme, an improved
version of Eq.(\ref{eq:taylor}) should be \cite[]{oldroyd:53}
\begin{equation}
  \label{eq:cm_taylor}
  \tilde\eta_T =\eta_c\frac{5+3(\eta_T/\eta_c-1)}{5-2(\eta_T/\eta_c-1)}\;.
\end{equation}
This equation predicts a larger viscosity than its truncation to first
order in $\phi$, Eq.(\ref{eq:taylor}). Both are shown as dot--dashed
lines in Fig.~\ref{fig:mixing}. Eq.(\ref{eq:cm_taylor}) is
\emph{qualitatively} superior to Eq.(\ref{eq:taylor}). We note,
however, that the limit $\lambda \to\infty$ deviates in second order
in $\phi$ from the result obtained for suspensions by
\cite{batchelor-green:72}.  Eq.(\ref{eq:cm_taylor}) and likewise all
of the following equations containing quantities $\tilde\eta_T$,
$\tilde\tau_1$, $\tilde\tau_2$ are only rigorous to first order in
$\phi$.

\begin{figure}
  \begin{center}
    \leavevmode 
     \includegraphics[width=0.55\columnwidth]{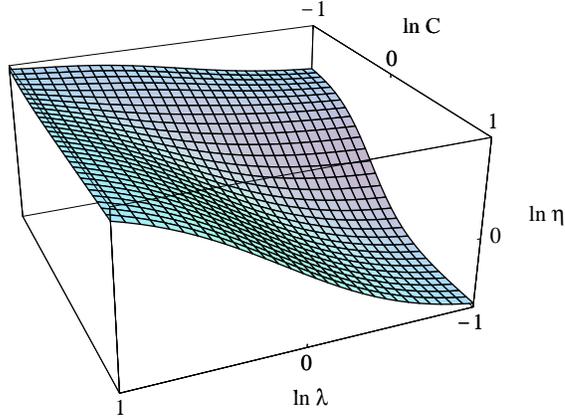}
     \caption{Eq.(\ref{eq:cm_miracle}) normalized to $\eta_c$ 
       as a function of viscosity ratio $\lambda$ and capillary number
       ${\cal C}$. The volume fraction of the dispersed phase is
       chosen to be $\phi=0.3$.}
    \label{fig:graph}
  \end{center}
\end{figure}

The same reasoning as to the viscosity applies to the characteristic
times $\tau_1$ and $\tau_2$ which now read \cite[]{oldroyd:53}
\begin{equation}
  \label{eq:tau_1_oldroyd}
  \tilde\tau_1=\frac{\eta_cR}{\sigma}
  \frac{[19\lambda+16][2\lambda+3-2\phi(\lambda-1)]}
  {40(\lambda+1)-8\phi(5\lambda+2)} \;,
\end{equation}
\begin{equation}
  \label{eq:tau_2_oldroyd}
  \tilde\tau_2=\frac{\eta_cR}{\sigma}
  \frac{[19\lambda+16][2\lambda+3+3\phi(\lambda-1)]}
  {40(\lambda+1)+12\phi(5\lambda+2)} \;.
\end{equation}
Their ratio $\tilde\tau_2/\tilde\tau_1$ is still given by
Eq.(\ref{eq:tau_2}).  Finally, Eq.(\ref{eq:miracle}) becomes
\begin{equation}
  \label{eq:cm_miracle}
  \begin{split}
    \eta  & = \tilde\eta_T
    \frac{1+\tilde\tau_1\tilde\tau_2\dot\gamma^2}{1+(\tilde\tau_1\dot\gamma)^2}
    \\
    & = \frac{\eta_c}{1+(\tilde\tau_1\dot\gamma)^2}
    \left(
      \frac{2\lambda+2+3\phi(\lambda+2/5)}{2\lambda+2-2\phi(\lambda+2/5)}+
      \frac{2\lambda+3+3\phi(\lambda-1)}{2\lambda+3-2\phi(\lambda-1)}
      (\tilde\tau_1\dot\gamma)^2
    \right) \;,
  \end{split}
\end{equation}
which to our knowledge has not been given before, and is one of our
main results. (For a graphical representation see
Fig.~\ref{fig:graph}.) In the limit $\tau_1\dot\gamma\to0$ it reduces
to Eq.(\ref{eq:cm_taylor}), whereas for $\tau_1\dot\gamma\to\infty$
only the second term in parentheses contributes and the \emph{curved}
dashed lines in Fig.~\ref{fig:mixing} are obtained. From the foregoing
discussion one should expect Eq.(\ref{eq:cm_miracle}) to be applicable
within a large range of shear rates, viscosities, and volume
fractions.

Finally, we note that more cumbersome expressions for $\tilde \eta_T$,
$\tilde \tau_1$ and $\tilde\tau_2$ have been derived within a different
cell model by \cite{choi-schowalter:75}. Here we only quote their
expression for $\tilde \eta_T$,
\begin{equation}
  \label{eq:choi}
  \frac{\tilde\eta_T}{\eta_c} \stackrel{C\&S}{=} 1+\phi
    \frac{2[(5\lambda)-5(\lambda-1)\phi^{7/3}]}
    {4(\lambda+1)-5(5\lambda+2)\phi+42\lambda\phi^{5/3}-5(5\lambda-2)\phi^{7/3}
    +4(\lambda-1)\phi^{10/3}}
\end{equation}
which is also represented graphically by the dotted lines in
Fig.~\ref{fig:mixing}. Since our data favor Eq.(\ref{eq:cm_taylor})
over Eq.(\ref{eq:choi}), and similar observations have been made by
others before (see Section~\ref{sec:experiment}), we will not pursue this
alternative approach further in the present contribution.

\subsection{A similarity rule}\label{sec:cox-merz}
It is interesting to observe that if morphology is conserved (drop
size $R$ independent of shear rate) for ${\cal C}\lsim {\cal C}^*$,
our Eq.(\ref{eq:cm_miracle}) for the nonlinear shear viscosity is
closely related to the expressions for the frequency dependent complex
viscosity $\eta^* (\omega)\equiv \eta' (\omega) +\text{i}\eta''(\omega)$ of
an emulsion of two incompressible Newtonian liquids as derived by
\cite{oldroyd:53},
\begin{equation}
  \label{eq:eta_w_oldroyd}
  \eta^* (\omega) = 
  \tilde\eta_T 
  \frac{1+\tilde\tau_2\text{i}\omega}{1+\tilde\tau_1\text{i}\omega} 
  \;, \qquad
  \eta' (\omega) = \tilde\eta_T
  \frac{1+\tilde\tau_1\tilde\tau_2\omega^2}{1+(\tilde\tau_1\omega)^2}
  \;, \qquad 
  \eta''(\omega) =  \tilde\eta_T
  \frac{(\tilde\tau_2-\tilde\tau_1)\omega}{1+(\tilde\tau_1\omega)^2}
  \;.
\end{equation}
For convenience, we also give the corresponding viscoelastic shear
modulus $G^*(\omega)\equiv \text{i}\omega\,\eta^*(\omega) \equiv
G'(\omega)+\text{i}G''(\omega)$,
\begin{equation}
  \label{eq:g_w_oldroyd}
  G'(\omega)= \omega\tilde\eta_T
  \frac{(\tilde\tau_1-\tilde\tau_2)\omega}{1+(\tilde\tau_1\omega)^2}
  \;, \qquad
  G''(\omega)= \omega\tilde\eta_T
  \frac{1+\tilde\tau_1\tilde\tau_2\omega^2}{1+(\tilde\tau_1\omega)^2} \;.
\end{equation} 
Obviously, the shear--rate dependent viscosity of
Eq.(\ref{eq:cm_miracle}) is obtained from the real part of the complex
frequency dependent viscosity $\eta^*(\omega)$ by substituting
$\dot\gamma$ for $\omega$,
\begin{equation}
  \label{eq:modified_cox-merz}
  \eta(\dot \gamma)\simeq \eta'(\omega)\;.
\end{equation}
In the same way, the first normal stress difference 
\begin{equation}
  \label{eq:fnsd_schowalter}
  p_{11}-p_{22}=-2\dot\gamma \cdot \phi\, \eta_c\, \tau_1\, \dot \gamma \,
  \frac{19\lambda+16}{(2\lambda+3)(2\lambda+2)}
\end{equation}
from Eq.(\ref{eq:schowalter}) is obtained to leading order in $\phi$
and $\dot\gamma$ from $2\omega\eta''(\omega)$, i.e.,
\begin{equation}
  \label{eq:first_normal_stress}
  p_{11}-p_{22} \simeq 2\omega\eta''(\omega)\;.
\end{equation}
Reverting the line of reasoning pursued so far, we can conclude that
to leading order in ${\cal C}$ and/or $\lambda^{-1}$ the weak
deformation limit of the constitutive equation of emulsions is
obtained from the linear viscoelastic spectra $G'(\omega)$, $G''
(\omega)$. Further, this identification can possibly be extended (at
least approximately) into regions of the ${\cal C}-\lambda$ plane
where the critical capillary number for drop breakup is somewhat
larger than ${\cal C}^*$ of Eq.(\ref{eq:region}).  In hindsight, it is
not surprising that in the case of weakly deformed drops the frequency
dependent viscosity and the steady shear viscosity are related. Note
that under steady shear, drops undergo oscillatory deformations at a
frequencey $2\omega=\dot\gamma$ if observed from a co--rotating frame
turning with vorticity at a frequency $\omega=\dot\gamma/2$. If we
take Eq.(\ref{eq:first_normal_stress}) seriously beyond the rigorously
known limit, we obtain an interesting prediction for the first normal
stress difference. In contrast to Eq.(\ref{eq:fnsd_schowalter}),
Eq.(\ref{eq:first_normal_stress}) implies that the first normal stress
difference saturates at a finite value
$40\,\phi\,\sigma/\!R[2\lambda+3+2\phi(1-\lambda)]^2$ for high shear
rates.  Thus, although the initial slope of the first normal stress
difference with $\dot \gamma$ increases with $\lambda$, its limit for
large $\dot \gamma$ decreases with $\lambda$.

Finally, we remark that based on qualitative theoretical arguments,
the similarity relation contained in Eq.(\ref{eq:modified_cox-merz})
and Eq.(\ref{eq:first_normal_stress}) has recently been proposed also
for polymer melts \cite[]{renardy:97}.  Usually, in polymer physics a
slightly different relation is considered; namely a similarity between
$\eta(\dot\gamma)$ and $|\eta^*(\omega)|$, also known as
\emph{Cox--Merz rule} \cite[]{cox-merz:58}.  In our case, since
$\eta''\!/\eta'=G'\!/G''=O(\phi)$, we can write
\begin{equation}
  \label{eq:cox-merz}
  \eta(\dot\gamma) \simeq |\eta^*(\omega)|+O(\phi^2) \;.
\end{equation}
Under the conditions mentioned at the beginning of this section, the
usual Cox--Merz rule is fulfilled to first order in $\phi$ for an
emulsion.  Eqs.(\ref{eq:modified_cox-merz}),
(\ref{eq:first_normal_stress}) are interesting from the theoretical
point of view, because they suggest a similarity of two \emph{a
  priory} rather different quantities. The results of this section
also can be of practical use, since they suggest that two different
methods may be applied to measure a quantity of interest.

\subsection{Non--Newtonian constituents}
Generalization of the above theoretical discussion to the case of
non--Newtonian constituents is not straightforward.  Indeed, as
\cite{oldroyd:53} already knew, his linear--response results quoted in
Eq.(\ref{eq:eta_w_oldroyd}) and Eq.(\ref{eq:g_w_oldroyd}) are readily
generalized to viscoelastic constituents by replacing the viscosities
$\eta_{d,c}$ in the expression for $\eta^*$ (or $G^*$) by complex
viscosities $\eta_{d,c}^* (\omega)$ \cite[]{palierne:90}. As a
consequence, the decompositions of $\eta^*$ and $G^*$ in real and
imaginary parts are no longer those of Eqs.(\ref{eq:eta_w_oldroyd})
and (\ref{eq:g_w_oldroyd}), and $\eta'$, $\eta''$, $G'$, $G''$ are
given by more cumbersome expressions.  For the steady--state
viscosity, on the other hand, one has to deal with a non--homogeneous
viscosity even within homogeneous regions of the emulsion, since the
strain rate itself is non--homogeneous and the viscosities are strain
rate dependent. We do not attempt to solve this problem here, nor do
we try to account for elasticity in the nonlinear case. Yet, it is an
intriguing question, whether the similarity rule
Eq.(\ref{eq:modified_cox-merz}) can be generalized to the case of
non-Newtonian constituents if the constituents themselves obey the
Cox--Merz rule (what many polymer melts and solutions do).  If both
constituents have similar phase angles $\theta\equiv \arctan G''\!/G'$
the generalized viscosity ratio
\begin{equation}
  \label{eq:general_lambda}
  \lambda^*\equiv \frac{\eta_d^*}{\eta_c^*} =
  \frac{|\eta_d^*|}{|\eta_c^*|}
  \text{e}^{\text{i}(\theta_d- \theta_c)} 
\end{equation}
that enters the expressions for $\eta^*$ and $G^*$, transforms
approximately to $\eta_d (\dot \gamma)/\eta_c (\dot \gamma)$ by
applying the Cox--Merz rule.  Therefore, in this particular example,
Eq.(\ref{eq:cm_miracle}) supports the expectation that the
generalization may work at least approximately.  If, on the other hand,
the phase angles of the constituents behave very differently, the
answer is less obvious.  This problem has been investigated
experimentally and is further discussed in
Section~\ref{sec:experiment}.

In any case, the generalization can only work if the representation of
the emulsion by a simple shear--rate dependent viscosity ratio $\eta_d
(\dot \gamma)/\eta_c (\dot \gamma)$, with $\dot\gamma$ the external
shear rate, is justified.  In the remainder of this section we
construct an argument that allows us to estimate the effective
shear--rate dependent viscosity ratio that should replace $\lambda$ in
Eq.(\ref{eq:cm_miracle}).  We take into account the deviation of the
strain rate from the externally imposed flow only within the drops,
because outside the drops the discrepancy is always small.  Inside a
drop, the strain rate can be small even for high external shear rates
if the viscosity ratio $\lambda=\eta_d/\eta_c$ is large.  Since we are
looking for an effective viscosity $\bar \eta_d (\dot \gamma)$ for the
whole drop to replace the viscosity $\eta_d$ at small shear rate, we
replace the non--Newtonian drop of non--homogeneous viscosity by an
effective pseudo--Newtonian drop of homogeneous but shear--rate
dependent viscosity. A possible ansatz for $\bar \eta_d$ is obtained
by requiring that the total energy dissipated within the drop remains
constant upon this substitution. Hence, we have
\begin{equation}
  \label{eq:dissipation}
  \int\!\!dV\; p_{ij}g_{ij} = 
  2\bar\eta_d \int\!\!dV\; \bar g_{ij}\bar g_{ij} \;,
\end{equation}
where $g_{ij}$ and $\bar g_{ij}$ denote the rate--of--strain fields in
the real drop and in the corresponding model drop of effective viscosity
$\bar \eta_d$, respectively. They both depend on the position within
the drop, whereas the average strain rate
\begin{equation}
  \label{eq:avg_strain}
  \dot\gamma^2_{\rm eff}\equiv \frac2V\int\!\!dV\; \bar g_{ij}\bar g_{ij}
\end{equation}
that enters the right--hand side of Eq.(\ref{eq:dissipation}) does
not.  Since the strain field $\bar g_{ij}$ within a drop of
homogeneous viscosity is unique for a given system in a given flow
field, $\bar \eta_d$ itself can be expressed as a function $\bar
\eta_d(\dot\gamma_{\rm eff})$ of the average strain rate.  Here, we
approximate this functional dependence by the strain rate dependence
$\eta_d(\dot\gamma)$ of the viscosity of the dispersed fluid.
Further, neglecting drop deformation we calculate $\dot\gamma_{\rm
  eff}$ from the velocity field within a spherical drop
\cite[]{bartok-mason:58} and obtain
\begin{equation}
  \label{eq:eta_eff}
  \bar \lambda\equiv \frac{\bar\eta_d(\dot\gamma_{\rm eff})}
  {\eta_c(\dot\gamma)} \approx 
  \frac{\eta_d\left(\sqrt2\dot\gamma/(\bar\lambda +1)\right)}
  {\eta_c(\dot\gamma)}\;.
\end{equation}
A different prefactor ($\sqrt7$ in place of $\sqrt2$) in the
expression for the effective strain rate $\dot\gamma_{\rm eff}$ was
obtained by \cite{debruin:89} using instead of the average in
Eq.(\ref{eq:avg_strain}) the maximum norm of $\bar g_{ij}$. To obtain
the correction to Eq.(\ref{eq:cm_miracle}) due to
Eq.(\ref{eq:eta_eff}) in the case of non--Newtonian constituents, the
implicit equation for $\bar \lambda$ has to be solved for given
functions $\eta_c(\dot \gamma)$ and $\eta_d(\dot\gamma)$. For shear
thinning constituents, Eq.(\ref{eq:eta_eff}) implies a tendency of
Eq.(\ref{eq:cm_miracle}) to overestimate $\eta(\dot \gamma)$ if
$\dot\gamma$ and $\lambda$ are large. In the actual case of interest,
for the constituents that were used in the experiments discussed in
Section~\ref{sec:experiment}, the viscosity ratio $\lambda$ ($\eta_d$
and $\eta_c$ taken at the external shear rate $\dot \gamma$) varies
almost by a factor of 10.  However, the corrections discussed in this
Section only become important for high shear rates, where the
constituents are shear thinning. In this regime, the viscosity ratio
(viscosities taken at the external shear rate) only varies between
$1/2$ and $2$, and hence the corrections expected from
Eq.(\ref{eq:eta_eff}) are at best marginally significant at the level
of accuracy of both Eq.(\ref{eq:eta_eff}) and the present
measurements. Therefore, a representation of the drops by
pseudo--Newtonian drops of homogeneous but shear--rate dependent
viscosity is most probably not a problem for the measurements
presented in the following section. The question as to a
generalization of the similarity rule Eq.(\ref{eq:modified_cox-merz})
to non--Newtonian constituents seems well defined.

\section{Experiment}\label{sec:experiment} 
\subsection{Materials and methods}
The experimental investigation deals with a phase separated aqueous
solution containing a polysaccharide (alginate) and a protein
(caseinate).  This type of solutions are currently used in the food
industry.  The methods for characterizing the individual polymers in
solution are in general known, especially when dealing with
non--gelling solutions where composition and temperature are the only
relevant parameters.  The polymers are water soluble.  When the two
biopolymers in solution are mixed, a miscibility region appears in the
low concentrations range and phase separation at higher
concentrations.  The binodal and the tie lines of the phase diagram
can then be established by measuring the composition of each phase at
a fixed temperature.  In general, the rheological behavior of phase
separated systems is difficult to investigate, and a suitable
procedure is not fully established.  In some cases, two--phase
solutions macroscopically separate by gravity within a short period of
time, but in some other cases (such as ours) they remain stable for
hours or days without appearance of any visible interface.  These
``emulsion type'' solutions have no added surfactant.  Following
approaches developed for immiscible blends, one may try to
characterize the partially separated solution as an effective emulsion
if the coarsening is slow enough. In order to establish a comparison
between phase separated solutions and emulsions, it is necessary to
know
\begin{itemize}
\item the volume fraction of the phases,
\item their shear--rate dependent viscosities (flow curves),
\item their viscoelastic spectra,
\item the interfacial tension $\sigma$ between the phases,
\item the average radius of $R$ the drops
\end{itemize}
Only the ratio $R/\sigma$ enters rheological equations. Knowledge of
either $R$ or $\sigma$ allows the other quantity to be inferred from
rheological measurements.

A difficulty when working with phase separated solutions, as opposed
to immiscible polymer melts, arises from the fact that each phase is
itself a mixture (and not a pure liquid) and therefore the rheology of
the phase depends on its particular composition.  If one wishes to
minimize the number of parameters, it is important to keep the
composition of the phases constant upon changing the volume fractions.
This can be achieved by working along a tie line of the phase diagram.
And this is precisely the procedure that we followed.  The polymers
were first dissolved, then a large quantity of the ternary mixture was
prepared (350 ml) and was centrifuged.  The two phases were then
collected separately. Both pure phases were found to be viscoelastic
and to exhibit shear thinning behavior, which is especially pronounced
for the alginate rich phase with $\eta(10^{-1}
\;\text{s}^{-1})/\eta(10^3 \;\text{s}^{-1})\approx 20$, while the
caseinate rich phase is almost Newtonian below $10^2$ s$^{-1}$.  The
viscosity of the alginate rich phase is higher than that of the
caseinate rich phase for shear rates below $2\cdot 10^2$ s$^{-1}$ and
lower for higher shear rates.  We checked that both phases obey the
usual Cox--Merz rule in the whole range of applied shear rates.

By mixing various amounts of each phase, the volume fraction of the
dispersed phase was varied between $10\,\%$ and $30\,\%$ while the
composition of each phase was kept constant.  In particular, the
temperature was kept constant and equal to the centrifugation
temperature in order to avoid redissolution of the constituents.  To
prepare the emulsion, the required quantities of each phase were mixed
in a vial and gently shaken.  Then the mixture was poured on the plate
of the rheometer (AR 1000 from TA Instruments fitted with a cone and
plate geometry 6 cm/$2^\circ$) and a constant shear rate was applied.
The apparent viscosity for a particular shear rate was then recorded
versus time until it reached a stable value.  By shearing at a fixed
shear rate, one may expect to create a steady size distribution of
droplets, with a shear rate dependent average size.  After each shear
experiment a complete dynamic spectrum was performed.  In this way,
shear rates ranging between $3\cdot 10^{-2}$ s$^{-1}$ and $10^3$
s$^{-1}$ were applied. The analysis of each spectrum according to
\cite{palierne:90} allowed us to derive by curve fitting the average
drop radius $R$ at the corresponding shear rate.

More technical details about the experimental investigation along with
more experimental results will be presented elsewhere.  Here, we
concentrate on the analysis of those aspects of the rheological
measurements pertinent to the theoretical discussion in
Section~\ref{sec:theory}.

\subsection{Results and discussion}
In this section we present our experimental observations and address
the questions posed at the end of the introduction. Before we present
our own data we want to comment briefly on related data recently
obtained for polymer melts by \cite{grizzuti-buonocore:98}.  These
authors measured the shear--rate dependent viscosity of binary polymer
melts and compared them to the low volume fraction limit of
Eq.(\ref{eq:cm_miracle}), i.e.\ Eq.(\ref{eq:miracle}), and to (a
truncated form of) results of \cite{choi-schowalter:75}. They reported
much better agreement with Eq.(\ref{eq:miracle}) than with the
truncated series from \cite{choi-schowalter:75}.  Comparison with the
full expressions of \cite{choi-schowalter:75} would have made the
disagreement even worse (cf.\ Fig~\ref{fig:mixing}). The average
radius $R$ that enters the equation, was determined independently for
each shear rate applied.  The constituents where moderately
non--Newtonian polymer melts, the viscosity ratio varying between
$\lambda\approx 0.3 \dots 3$ over the range of shear rates applied.
Hence, these experiments, are located in the interesting parameter
range, where Eqs.(\ref{eq:cm_miracle}) and Eq.(\ref{eq:miracle}) for
$\eta (\dot \gamma)$ are expected to be sensitive to drop deformation
and break--up. Surprisingly, the results show that they describe the
data very well over the whole range of shear rates although one would
not necessarily expect average drop deformation to be very small.
Unfortunately, drop sizes have not been reported by the authors, so
conclusions concerning the location in the ${\cal C}-\lambda$
parameter plane and the validity of the similarity rule
Eq.(\ref{eq:modified_cox-merz}) cannot be drawn. Also the question,
whether Eq.(\ref{eq:cm_miracle}) holds for small viscosity ratios
$\lambda \ll 0.3$, cannot be answered.

\begin{figure}
  \begin{center}
    \leavevmode 
     \includegraphics[width=0.8\columnwidth]{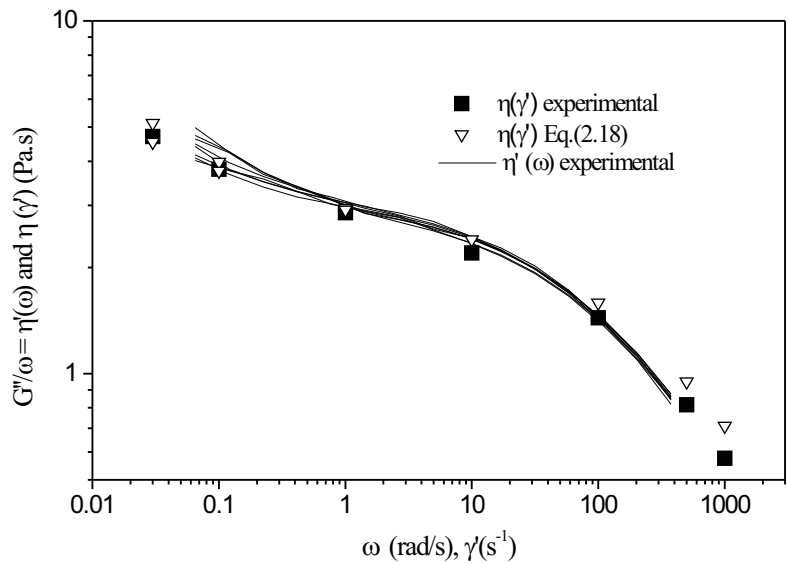}
     \caption{The nonlinear shear viscosity $\eta (\dot\gamma)$
       (opaque squares) and the real part $\eta' (\omega)$ of the
       dynamic viscosity $\eta^* (\omega)$ (lines) of a droplet phase
       of a mixture of weakly viscoelastic polymer solutions
       (alginate/caseinate). Also shown is Eq.(\ref{eq:cm_miracle})
       for the viscosity of an emulsion of Newtonian constituents
       evaluated for the actually non--Newtonian viscosities of the
       constituting phases with the drop size obtained from the
       spectra (open triangles). Due to the scatter in the dynamic
       viscosity at low frequencies there is an uncertainty in the
       average drop size, resulting in corresponding error bars
       (multiple points) for Eq.(\ref{eq:cm_miracle}).}
    \label{fig:cox-merz}
  \end{center}
\end{figure}

Our own measurements were located in about the same $\lambda-$range.
As we noted in the preceding section, only the ratio $R/\sigma$ enters
rheological equations and knowledge of either $R$ or $\sigma$ allows
the other quantity to be inferred from rheological measurements.
\cite{ding-pacek:99} determined the interfacial tension of the
alginate/caseinate system used in our experiments by observing drop
relaxation under a microscope and analyzing the data according to
Section~\ref{sec:theory}. They found $\sigma \simeq 10^{-5}$ N/m.
Using this, we obtained an average drop size $R\simeq 10^{-5}$ m from
the measured spectra $G'(\omega)$, $G''(\omega)$ according to
\cite{palierne:90} for the experiments reported in
Fig.~\ref{fig:cox-merz}.  By the method based on \cite{palierne:90},
we could not detect a decrease in drop size with shear rate as
expected from the phenomenological phase diagram for single Newtonian
drops under shear as established by \cite{grace:82} and others. Thus,
the limit of high capillary numbers and moderate viscosity ratios (the
region above the break--up curve) in Fig.~\ref{fig:limits} has been
accessed experimentally.  Corresponding locations have been indicated
qualitatively in the figure by dashed lines. With the drop size being
constant, one can try to test the proposed similarity rule
Eq.(\ref{eq:modified_cox-merz}).  By identifying the axis for
frequency $\omega$ and shear rate $\dot\gamma$, data for the real part
$\eta' (\omega)$ of the frequency dependent dynamic viscosity $\eta^*
(\omega)$ are compared to data for the shear--rate dependent viscosity
$\eta (\dot \gamma)$ in Fig.~\ref{fig:cox-merz}.  The emulsion
containing $30\,\%$ of the alginate rich phase and $70\,\%$ of the
caseinate rich phase has been prepared at room temperature as
described in the preceding section.  Steady shear rates ranging
between $3\cdot 10^{-2}$ s$^{-1}$ and $10^3$ s$^{-1}$ corresponding to
capillary numbers ${\cal C}\approx 6\cdot 10^{-2} \dots 10^{3}$ have
been applied. The shear viscosity (opaque squares) is reported in the
figure for each of these individual measurements.  The multiple data
sets for $\eta'$ (lines) taken each between two successive steady
shear measurements, superimpose fairly well; i.e.\ the spectra appear
to be remarkably independent of the preceding steady shear rate. The
good coincidence of $\eta'(\omega)$ and $\eta (\dot \gamma)$ in
Fig.~\ref{fig:cox-merz} show that the data obey the proposed
similarity rule Eq.(\ref{eq:modified_cox-merz}) over a large range of
shear rates. The agreement near $\dot\gamma\approx 2\cdot 10^2$
s$^{-1}$ is a consequence of the proximity to the trivial limit
$\lambda=1$, ${\cal C}=\infty$.  Nevertheless, the data provide strong
evidence that Eq.(\ref{eq:modified_cox-merz}) is an excellent
approximation for a large range of viscosity ratios and capillary
numbers. Similar results (not shown) have been obtained for other
volume fractions.  Comparison with Eq.(\ref{eq:cm_miracle})
represented by the open triangles in Fig.~\ref{fig:cox-merz}, on the
other hand, is less successful at large shear rates, although it is
still not too far off for a theoretical curve without any adjustable
parameter. A discrepancy had to be expected as a consequence of the
non--Newtonian character of the constituents at large shear rates,
which is definitely not taken into account in
Eq.(\ref{eq:cm_miracle}). For the plot of Eq.(\ref{eq:cm_miracle}) in
Fig.~\ref{fig:cox-merz} we merely substituted $\eta_d
(\dot\gamma)/\eta_c (\dot\gamma)$ taken at the external shear rate
$\dot\gamma$ for $\lambda$. The average drop size $R$ was obtained
from fitting the viscoelastic moduli. The scatter in the dynamic
viscosity data gives rise to an uncertainty in $R$, which is reflected
by the multiple open triangles at low shear rates. It seems that the
similarity rule Eq.(\ref{eq:modified_cox-merz}) is more general than
Eq.(\ref{eq:cm_miracle}), i.e., it still holds for rather viscoelastic
constituents (that obey the usual Cox--Merz rule), where the latter
fails. This relation certainly deserves further investigation with
different materials and methods.

In summary, we have succeeded in establishing an analogy first between
a partially phase--separated polymer solution and an emulsion, and
further between the viscoelastic spectrum of the system and its
nonlinear shear viscosity even in the case of (moderately)
non--Newtonian constituents.

This work was supported by the European Community under contract no
FAIR/CT97-3022. We thank P. Ding and A. W. Pacek (University of
Birmingham) for measuring the surface tension and S. Costeux and G.
Haagh for helpful discussions and suggestions.

\bibliographystyle{jfm}

\end{document}